# Low-Loss Higher-Order Cross-Sectional Lamé Mode SAW Devices in 10-20 GHz Range

Ian Anderson[1], Tzu-Hsuan Tsu[1,2], Vakhtang Chulukhadze[1], Jack Kramer[1], Sinwoo Cho[1], Omar A. Barrera[1], Joshua Campbell[1], Ming-Huang Li[2], and Ruochen Lu[1]

[1]University of Texas at Austin, Austin, TX, USA, [2]National Tsing Hua University, Hsinchu, Taiwan, Email: ianderson@utexas.edu

*Abstract*—This paper presents surface acoustic wave (SAW) acoustic delay lines (ADL) for studying propagation loss mechanisms in Lithium Niobate (LN). Devices were fabricated by depositing 50 nm aluminum patterns on 600 nm X-Cut LN on amorphous silicon on silicon carbide, where longitudinally dominant SAW was targeted. Upon fabrication, higher-order thickness-based cross-sectional Lamé modes and Rayleigh modes were studied for their Q factors using acoustic delay lines. Utilizing bi-directional electrodes, ADL with lateral lambda ($\lambda$) values ranging from 0.4 μm to 0.6 μm were measured. Higher order Lamé modes were found to have consistently higher Q factors than their Rayleigh mode counterpart, on the order of 1000-3000, showing high-frequency SAW devices as still viable candidates for frequency scaling without a substantial increase in loss.

## I. Introduction

Acoustic wave-based devices, used for signal processing technologies such as filters or delay lines, have been the leading solutions as compact filters for the cellular industry since their inception [1], [2]. Acoustic waves feature lower loss and lower size as compared to electromagnetic (EM) waves, making acoustic based devices the ideal solution for mobile phones where size and power requirements are incredibly stringent [3], [4]. Converting EM signals to acoustic signals requires piezoelectric materials and transducer geometry that scales with device frequency. While this geometry scaling proved advantageous on the micron scale, as 6G technologies target higher and higher frequency bands, the fabrication of transducers becomes tedious and inconsistent at the nanometer scale [5], [6]. To scale surface acoustic wave (SAW) based devices to higher frequencies, one solution involves using higher velocity modes, such as the longitudinal SAW (LSAW) wave [7]-[10]. Because this acoustic wave is based largely on the much faster longitudinal wave velocity in the material, it allows devices to be fabricated at much higher frequencies using the same lateral dimensions [11]. In addition, with advancements in fabrication technology, the prospect of fabricating higher and higher frequency SAW devices by simply scaling down dimensions seems to be on the horizon as well [12]. Combining these two ideas, it seems that high-frequency SAW is feasible in the near future, though little is known about their intrinsic loss mechanisms. Many material models suggest loss mechanisms scaling with frequency,

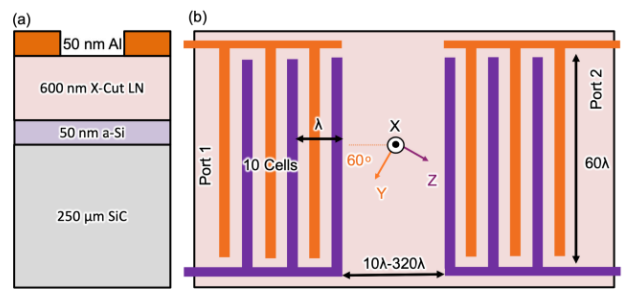

Figure 1: (a) Stack layer information and (b) Topology of acoustic delay line testbed using bidirectional electrodes.

threatening to once again put a limit on device operation frequency [13]-[16].

This study presents generalized cross-sectional Lamé modes for 10-20 GHz operation in thin-film X-Cut LN on silicon carbide (SiC). We used acoustic delay lines (ADLs) as testbeds to showcase acoustic wave excitation and propagation. We report a method for calculating propagation losses in ADLs to calculate the quality factor ($Q$) of each mode [17]. ADLs with identical transducers but varying delay lengths are implemented to extract propagation loss in materials. This method for determining the $Q$ is favorable over other theoretical models in that it allows us to avoid other loss mechanisms such as electrical feedthrough [18], [19]. High $Q$s of around 2000 are reported for cross-sectional Lamé modes between 10 and 20 GHz (Fig. 6), achieving remarkable frequency scaling without increasing loss with larger Q than corresponding Rayleigh modes utilizing thickness-based modes on SAW platforms.

## II. Device Design and Simulation

Simulation of 600 nm X-Cut LN on 50 nm amorphous silicon on 4H SiC stack is seen in Fig. 1 featuring simulation of ADL and eigenmode analysis of corresponding modes utilizing Finite Element Analysis (FEA) via COMSOL Multiphysics. LN is used for strong piezoelectric coupling coefficients, while SiC is used for rigid interface for wave confinement, thanks to its high velocity and stiffness [20]-[22]. Perfectly matched layers (PML) are used at either end of the ADL, as well as on the bottom side of the SiC to avoid reflected waves, representing the bulky substrates. Devices are rotated 60° in-plane to maximize the $e_{11}$ and $e_{35}$ coefficients for excitation of proposed modes. Simulations of device performance are shown in Fig. 2 for a device with λ of 0.4 μm and gap length of 40λ. Among the mode types includes the low-frequency Rayleigh mode at 7.6 GHz, along with several higher thickness orders of the cross-sectional Lamé modes at 10.1 GHz, 12.1 GHz, 14.7 GHz and 16.8 GHz. These modes appear alongside our expected LSAW mode, which is found at 13.5 GHz, originating from the copropagating shear vertical wave that gives rise to thickness modes due to the strong reflection from LN and the highly rigid SiC interface.

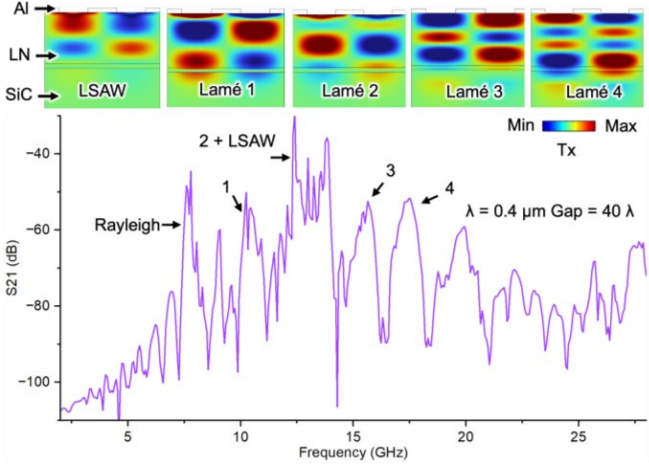

Figure 2 Simulation data showing Rayleigh mode, LSAW mode, and higher order thickness lamé modes from 10-20 GHz.

Because the device now has two wave vector components contributing to wave propagation, our estimation for center frequency can be determined by the below (1), where $v_l$ and $v_t$ are the transverse and longitudinal wave speeds, $t_{LN}$ is the thickness of LN, $\lambda$ is still our lateral wavelength, and n is our thickness component order. Our increase in mode order n corresponds to an increase in frequency of each passband. This allows our modes to increase in frequency without changing lateral dimensions, all on solidly mounted platforms. This equation aligns with the increase in frequency of simulations, with all values being off by about 1 GHz.

$$f_c \approx \sqrt{\left(\frac{v_l}{2\lambda}\right)^2 + \left(\frac{n \cdot v_t}{2 t_{LN}}\right)^2} \quad (1)$$

Also shown in Fig. 2 is the x-component of stress for the first four Lamé modes and the LSAW mode, showing the same lateral lambda but increasing thickness order with each passband for Lamé modes. While these modes are longitudinally dominant in the x component of stress, they also contain z-components, hence higher order thickness modes.

Fig. 3 shows dispersion curves for the devices over a range of $t_{ln}/\lambda$ ratio compared to frequency and wave velocity ($v_p$) for a constant thickness of $t_{LN}$ = 600 nm while varying $\lambda$. As the ratio between the two gets larger, approaching higher frequency, higher order Lamé modes pop up increasingly, with our third and fourth order Lamé mode appearing around $t_{ln}/\lambda$ = 1.0 and 1.3 respectively ($\lambda$ = 0.46 μm and 0.58 μm). These modes disappear once their wave velocities become larger than that of the shear vertical wave in SiC at around $v_p$ = 7100 m/s, and the modes become too lossy [23]. It is also noted that higher-order modes all have much larger phase velocity around 6000 m/s, compared to 3500 m/s Rayleigh mode speed, and all modes exhibit a decrease in phase velocity with increasing $t_{ln}/\lambda$ (decrease in $\lambda$). The first order Lamé mode has cutoff around $t_{ln}/\lambda$ = 0.24, not shown on graph, while LSAW mode has no cutoff frequency, as it is a surface wave not reliant on rigid SiC boundary for reflection of shear vertical wave.

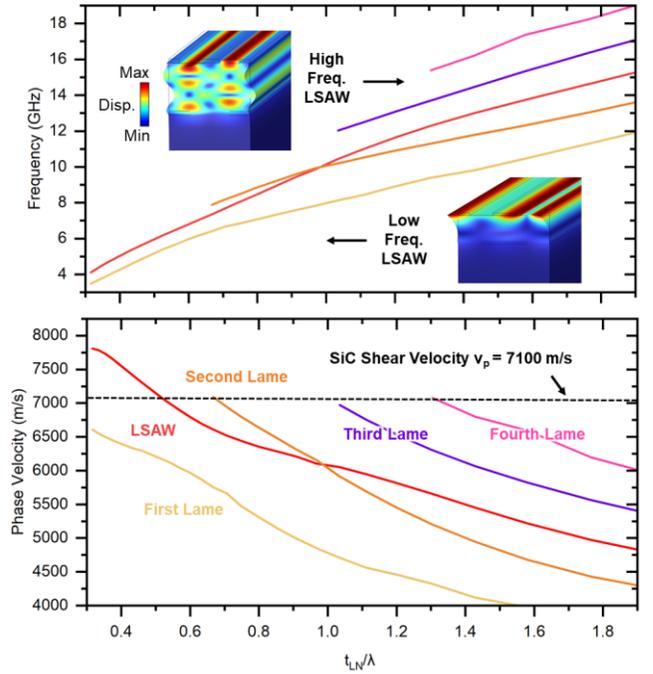

Figure 3: Dispersion curves for various modes showing frequency and phase velocity, highlighting where they cutoff for specific values of $t_{LN}/\lambda$.

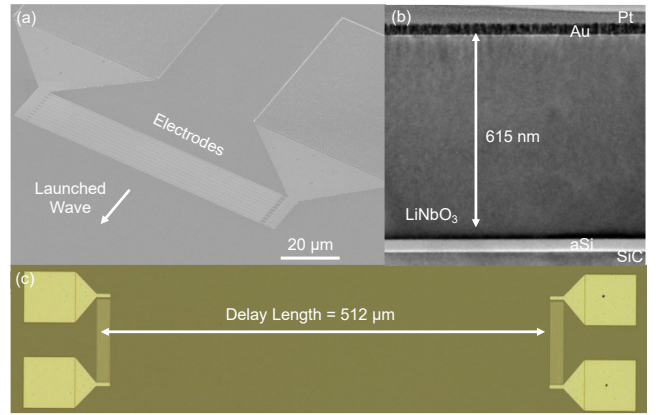

Figure 4: (a) SEM, (b) TEM, and (c) optical images of acoustic delay lines, showing topology and sideview profile.

In addition, the LSAW mode crosses the second Lamé mode around a ratio of 1, and becomes a Lamé-like mode, gaining thickness elements just as the other higher-order modes do. In Fig. 3, low frequency LSAW appears as normal, while high frequency LSAW appears more like a Lamé mode. At around $\lambda$ = 0.4 μm, the second-order Lamé mode and the LSAW mode have overlapping frequencies, leading to the very uneven passband at 13 GHz.

III. DEVICE MEASUREMENT

Devices are fabricated via E-Beam lithography and evaporation of 50 nm thick aluminum electrodes and bus line. SEM and optical images of fabricated devices are seen in Fig. 4, featuring a TEM image showing our X-Cut LN with a thickness of 615 nm, an SEM of the transducer topology, and an optical image of an entire device. ADLs were fabricated at several lateral wavelength (λ) values from 0.4-0.6 μm with several delay length values from 10-320λ. Other ADL parameters include the 10 cells, corresponding to a total of 20 electrodes on each side, the 60λ aperture width (seen in Fig. 1), and a bi-directional electrode configuration featuring λ/4 electrode widths and gaps.

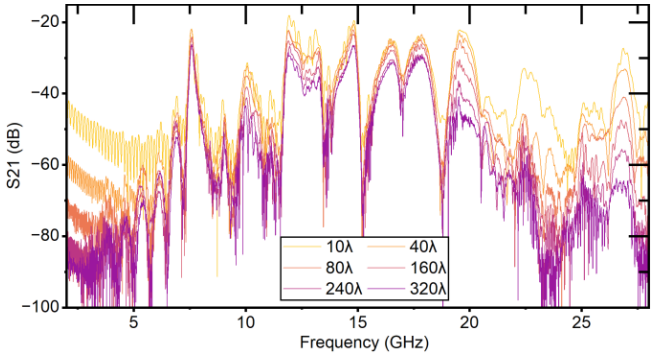

Figure 5: Example S21 data from 6 devices of the same λ but varying gap lengths, showing a drop in insertion loss for each mode.

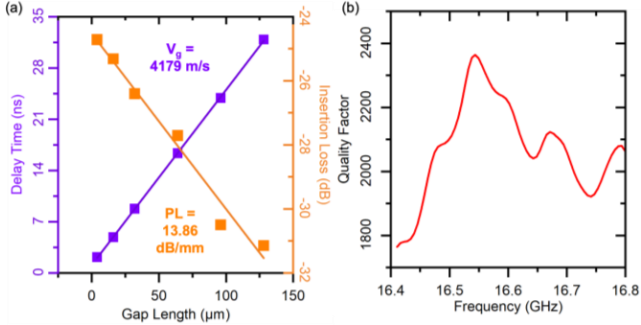

Figure 6: (a) shows example of gradients in insertion loss and group delay used to calculate and (b) example Q over passband for third Lamé mode in the λ = 0.4 μm device.

Devices are measured using a Keysight P5027A network analyzer at -15 dBm power level. Measurement results of transmission $S_{21}$ from devices can then be seen in Fig. 5, which displays data from the same lateral lambda devices with varying delay lengths. Results seen here consist of data after time gating, a process where the signal is isolated in the time domain, and the EM signal is removed to show higher prominence of acoustic modes as seen in [24]. Measured data matches simulated data well, with passband for Rayleigh mode at 7.8 GHz, and higher order Lame/LSAW Modes with passbands at 10 GHz, 12 GHz, 14.7 GHz, 16.2 GHz, and 19.5 GHz. Important to note of these devices is the increase in loss between the devices with increasing delay lengths. Because delay length is the only difference between the devices, we know that the extra loss seen from each subsequent device is from propagation loss, which allows us to derive the $Q$ of the material and mode from (2) below:

$$Q = \frac{f \cdot \pi}{PL \cdot v_g} \quad (2)$$

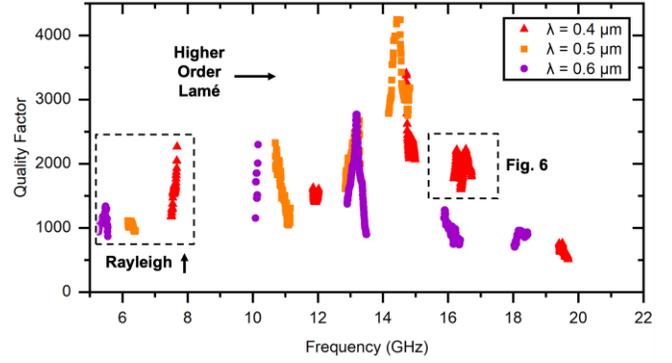

Figure 7: Q over passband for each device, showing lower Q for Rayleigh modes and tapering Q at high frequency for Lamé modes.

TABLE I COMPARISON TO SOA

| Ref. | Platform | Solidly Mounted | f (GHz) | Q |
|---|---|---|---|---|
| S. Cho, et. al. [17] | ScAlN | No | 5-10 | ~600 |
| J. Kramer, et. al. [24] | P3F LiNbO$_3$ | No | 10-30 | ~1000 |
| C. Tsai, et. al. [25] | LiNbO$_3$ on SiO$_2$/Si | Yes | 0.9 | ~1500 |
| G. Giribaldi, et. al. [26] | ScAlN | No | 8-10 | ~2000 |
| C. Yeh, et. al. [27] | LiNbO$_3$ on SiO$_2$/Si | Yes | 0.9 | ~675 |
| **This Work** | **LN on SiC** | **Yes** | **10-20** | **~2000** |

where PL is the propagation loss (dB/mm), derived from the gradient of insertion loss in Fig. 6, f is the frequency and $v_g$ (m/s) is the group velocity, obtained from the gradient of the group delay of each device, also shown in Fig. 5. Examples of the derivation of each of these values (PL and $v_g$) are shown in Fig. 6(a), where the insertion loss and group delay are plotted for each device, and a linear regression model is fitted to the data to extract the parameters. Large variation in the Q of each device throughout the passband is noted, particularly for high-frequency Lamé modes, due to the in-band ripples of each device. Reflections from one transducer to the other lead to oscillations in the insertion loss within the passband, leading to oscillations in the $Q$ factor.

With these parameters and the frequency of each device, Q factors are calculated over the passband of each device and are shown in Fig. 7, with a zoom-in on the passband of the third Lamé mode in the λ = 0.4 μm device in Fig. 5(c). While our higher-order Lamé modes exhibit more propagation loss, as can be seen from Fig. 5, their increase in frequency from the Rayleigh modes gives them larger Q values. From this, we can see that our higher frequency pass bands of the Lamé mode show consistently high $Q$, around 2000 while Rayleigh modes exhibit lower Q values, around 1000 for λ = 0.5 μm and 0.6 μm or 1750 for λ = 0.4 μm.

Comparison to other studies of Q factor shows our ADLs have consistently higher Q factors at higher frequency (Table I). These high material and mode Q values for Lamé modes highlight their potential use for high frequency SAW in the future.

## IV. Conclusion

This work explores loss values in high order SAW waves utilizing ADL in X-Cut LN on SiC. Designs were targeted at LSAW modes, which produced several higher order modes at 10-20 GHz due to the copropagating of shear vertical waves with LSAW mode. Q values of 1000-3000 were found for said high order Lamé modes, displaying the feasibility of scaling SAW waves toward higher frequency bands.

## Acknowledgment

The authors thank the DARPA COFFEE program for funding support and Dr. Ben Griffin and Dr. Todd Bauer for helpful discussions.